\documentstyle[preprint,aps]{revtex}
\begin{document}
\draft

\title{Motional diminishing of optical activity: a novel method for studying 
molecular dynamics in liquids and plastic crystals}

\author{Michael C. Martin\thanks{Current address:  Deptartment of Physics, 
Building 510B, Brookhaven National Laboratory, Upton, NY  11973-5000} 
and Laszlo Mihaly}
\address{Department of Physics, State University of New York at Stony Brook, 
Stony Brook, NY 11794-3800}

\date{\it To be published in Chemical Physics Letters.}
\maketitle

\begin{abstract}
Molecular dynamics calculations and optical spectroscopy measurements 
of weakly active infrared modes are reported.  
The results are qualitatively understood in terms of the 
``motional diminishing" of IR lines, a
process analogous to the motional narrowing 
of a nuclear magnetic resonance (NMR) signal.  In molecular solids or 
liquids where the appropriate intramolecular resonances are observable, 
motional diminishing can be used to study the fluctuations of the 
intermolecular interactions having time scales of 1psec to 100psec.  
\end{abstract}

\newpage
\narrowtext

Group theory of molecular symmetries is extensively used in infrared and 
Raman spectroscopies to assign vibrational modes to point group 
representations.  One of the simplest and most important arguments applies 
to molecules with a center of inversion symmetry:  the vibrations can 
be divided into symmetric ({\it gerade}, denoted $g$) and asymmetric 
({\it ungerade}, denoted $u$) modes; only a $u$ mode may have IR-activity 
and only a $g$ mode is allowed in the Raman spectrum.  Except for the 
simplest molecules, further analysis is needed to determine if the modes 
are indeed allowed; the $u$ modes with other than vector symmetry are 
IR-forbidden,  and the $g$ modes with other than tensor symmetry are Raman 
forbidden.  

The symmetry group of an isolated molecule is strictly relevant only for low 
density gas phase samples.  Intermolecular interactions break the pure 
molecular symmetry.  In the gas phase, ``collision induced scattering" 
(CIS) has been observed for some time, leading to an enhanced 
Rayleigh tail in noble gases \cite{argon,argon2}, a broadening of the 
Raman line in SF$_6$ \cite{sf6} and the appearance of ``forbidden" IR modes 
in the Raman spectrum of CO$_2$ \cite{co2}.  In liquids the intermolecular 
forces are even stronger and a large body of literature discusses the 
various manifestations of these interactions 
\cite{jonas1,birnbaum,oxtoby,tabisz2,maddencox}.  In the solid state 
the static perturbation leads to the well 
known crystal field splittings of degenerate modes, in addition to the 
appearance of previously ``forbidden" resonances in the spectrum 
\cite{ramansolidcs2,ramansolidcs22,irsolidcs2,irsolidcs22,irsolidc60}.  

For disordered systems the treatment often starts in terms of 
two experimental time scales: the energy relaxation time $T_1$ and the 
dephasing time $T_2$ \cite{oxtoby,lynden}.  These times are analogous to 
$T_1$ and $T_2$ in nuclear magnetic resonance (NMR)\cite{slichter}.  
$T_1$ tells us 
how fast the energy of a given vibration is dissipated to other modes; $T_2$ 
is the time required for the oscillators in the ensemble to develop a $2\pi$ 
phase difference.  Since in many systems $T_1\gg T_2$, the width of a 
resonance line is, in good approximation, $\Delta \omega = 1/T_2$.  

The theory of the gas and the solid phases are relatively easy to handle, 
whereas the liquid presents a greater challenge.  For CIS in the gas phase 
the dephasing is viewed as a result of independent collision processes, 
separated by the average time between collisions $\tau_c$.  Each collision 
adds a small random phase to the oscillations; consequently, 
$\Delta \omega \sim 1/\tau_c$.
The magnitude of phase shift depends on the ``violence" of the collision 
relative to the strength of the oscillator.  A detailed calculation 
results in $\Delta \omega=(k_{\rm B}T)/(\omega^2 L^2 m \tau_c$), where 
$k_{\rm B}T$ is the thermal energy, $L$ is the range of interaction and 
$m$ is a mass factor \cite{oxtoby2}.  In the crystalline solid 
state the problem is formulated in terms of the exchange coupling energy 
$V$ between the oscillators: the line splittings will be of the order of 
$\delta \omega \propto V$.  In a disordered solid the a linewidth $\Delta 
\omega \propto <V^2>^{1/2}$ is expected, 
where $<>$ represents an appropriate average over all configurations.  

Approaching from these two extremes, the liquid can be either viewed as a 
dense gas, or as a disordered, fluctuating solid.  A great amount effort has 
been concentrated on adapting the gas phase results to liquids 
\cite{oxtoby2}, including calculations taking into account the microscopic 
details of the interaction potential \cite{maddencox,coxmadden}.  In this 
Communication we will take the other route.  We will investigate what happens 
to the 
linewidth of the allowed resonance lines, and to the width and strength of 
the ``forbidden" lines, if fluctuations are introduced to a quasi-static, 
random system.  We perform computer simulations on a classical 
vibrational model and we use CS$_2$ as a model system in the experiments.  
The agreement we find between the measurement and the calculations 
should not be viewed as an exhaustive and quantitative explanation for the 
behavior of the spectral lines, however the results do point to a novel 
contributing mechanism, which we term ``motional diminishing".  

Qualitatively, as the correlation time of the fluctuations ($\tau $) 
decreases, one expects that the effective symmetry breaking field will 
disappear.  This is evident in the limit of extremely fast fluctuations, 
when $\tau <T$ (where $T=2\pi/\omega $, the period of oscillation of the 
molecular vibration in question) the external fields are expected to 
completely average out, producing effectively no symmetry breaking.  
Consequently, $i.$ the linewidth of the allowed resonances narrows, 
$\Delta \omega \propto \tau$ \cite{lynden}, and 
$ii.$ the weakly allowed lines disappear.  The question is, what is 
the critical timescale of the fluctuations for this crossover to occur?  
While the first effect (``motional narrowing") has been discussed 
before \cite{oxtoby,lynden}, the crossover from weak optical activity 
to no activity as the correlation time of the fluctuations decreases is a 
novel phenomena; we call it motional diminishing.  

In addition to shedding a new light on the behavior of molecular liquids, 
our considerations are directly applicable to ``plastic" crystals 
\cite{winsor}.  In fact the main motivation of 
this work originated from the desire to understand the behavior of a new and
particularly interesting plastic crystal, C$_{60}$ \cite{c60,c602,c603}.  
In a plastic crystal the 
molecules perform rotational diffusion, while preserving their center of mass 
positions at the crystallographic coordinates.  If viewed from the 
reference frame of a given molecule, the {\em magnitude} of the crystal 
field does not change much but its {\em direction} seems to fluctuate with 
a correlation time $\tau$, the rotational diffusion time constant.  
Note that $\tau$ and
$\tau_c$ from CIS have opposite effects:  $\Delta \omega \propto \tau$ and
$\Delta \omega \propto 1/\tau_c$

We performed computer simulations by solving the classical equations of 
motion for a simple, one dimensional, harmonic model system, in the presence 
of a harmonic driving field E of angular frequency $\omega$, 
and fluctuating external perturbations.  The time 
averaged power dissipation was calculated by integrating the 
field$\times$charge$\times$velocity at various values of the drive frequency.  
The ``ball and spring" system is similar to a CS$_2$ or 
CO$_2$ molecule (except that it does not model the bending mode): it 
consists of two balls of equal mass $m$ and 
charge $q$ attached to a third ball of mass $M$ and charge $Q$ by equivalent 
springs of spring constant $k$.    
This ``molecule" has a center of inversion symmetry with 
one Raman-active $g$ mode and one IR-active $u$ mode.  The external 
symmetry breaking perturbation 
is represented by two springs, $k^{\prime}$ and $k^{\prime \prime}$, 
connected to the two outer masses.  The Hamiltonian of this system is,
\begin{eqnarray}
&H={p_1^2\over{2m}}+{p_2^2\over{2m}}+{p_3^2\over{2M}}+{1\over{2}}k(x_1-x_3)^2
+{1\over{2}}k(x_2-x_3)^2 \nonumber \\
&+E[q(x_1+x_2)+Qx_3]+{1\over{2}}k^{\prime}x_1^2
+{1\over{2}}k^{\prime \prime}x_2^2\ .
\end{eqnarray}
Randomness of the external perturbation 
is achieved by making the strengths of these outer springs fluctuate 
randomly.  The fluctuations are characterized by a correlation time $\tau $; 
the probability of switching to a new random value of $k^{\prime}$ (or 
$k^{\prime \prime}$) in time interval d$t$ is d$p=$d$t/\tau $.  All 
other effects of the environment, the coupling to other modes, 
{\it etc.}, are represented by a damping force proportional 
to the velocity of the masses, with damping constant $\gamma $ for all three 
balls.  This damping provides a finite width for each resonance mode even 
if no outside perturbation is provided.  Typical numerical values were 
$k=1$, $m=M=1$, $q=-1$, $Q=+2$ and $\gamma=0.005$.

If one side of the system is perturbed by an time-independent 
potential (represented by $k^{\prime}$), then two resonance lines are 
observed: a weak one close to the ``forbidden" $g$ mode frequency 
$\omega=\omega_0=\sqrt{k/m}$ and a strong one around the $u$ mode, 
$\omega=\sqrt{3}\omega_0$.  The line shifts from the ``ideal" 
positions are linear in $\kappa=k^{\prime}/k$; the line intensity of the 
``forbidden" resonance is proportional to $\kappa^2$.

First we consider random fluctuations where a perturbing spring of 
fixed magnitude is randomly switched between the 
left and right sides, the unperturbed side having zero external coupling.  
In this system (referred to as ``case I" later), the resonance mode 
frequency is the same for both the ``left" and ``right" perturbations, and 
therefore there is no additional line broadening due to the external forces.
We find that slowly fluctuating perturbations produce 
a weak IR mode close to $\omega=\omega_0$, but this ``forbidden" resonance 
vanishes as the 
correlation time $\tau$ becomes less than the lifetime of the resonance 
$\tau<T_1=1/\gamma$.  Note that the line narrows even if $\tau$ is much 
longer than the period of the oscillation.

A more realistic model should involve random fluctuations in the magnitude 
of the external perturbations (this will be case II).  Figures \ref{gmode} 
and \ref{umode} summarize the results obtained when the $k^{\prime}$ and 
$k^{\prime \prime}$ spring constants were assumed to vary independently and 
randomly with a uniform probability distribution in the range of 
$0<\kappa<\kappa_{max}=0.4$.  For quasi-static perturbations (long 
correlation times such as $\tau=5000$) the $u$ mode and the $g$ mode 
resonances are both broadened beyond the lifetime effects seen in case I
(the quasi-static linewidth $\Delta \omega (\tau \rightarrow \infty)=\delta$ 
is larger than $\gamma$), and the center frequencies are shifted from the 
unperturbed frequencies to somewhat higher values.  
At the other time extreme, fast fluctuations (like $\tau=5$) lead to a 
vanishing infrared activity of the $g$ mode (Fig. \ref{gmode}).  
The $u$ mode narrows to the ``lifetime" width $\gamma$, 
and its integrated intensity remains approximately unchanged
(see the upper inset of Figure \ref{umode}).  

The parameters in this work are deliberately selected so that the three time 
scales in the problem ($T=2\pi /\omega$,  $T^{\star}\equiv 1/\delta$, and 
$T_1=1/\gamma$) are well separated.  For the $g$ mode, in the 
dimensionless units of the calculation, $T=1.102$, $T^{\star}=18$ and 
$T_1=200$; for the $u$ mode the values are 1.758, 40 and 
200, respectively.  The insets in Figures \ref{umode} and \ref{gmode} 
clearly indicate that the crossover from the quasi-static behavior at long
$\tau$ (broadened IR line and non-zero IR intensity of the Raman line) to 
the fast fluctuations regime (narrowed IR line and vanishing IR-activity 
of the Raman line) occurs at around $\tau=T^{\star}$ for each mode.  The
intensity of the ``forbidden" Raman line as a function of $\tau$ is well 
fit by the function $I=I_0/(1+{T\over{\tau }})$, as demonstrated by the 
solid line in the inset to Figure \ref{gmode}.  In contrast to case I, the 
lifetime of the resonance, $T_1$, does not play a particularly 
important role in the physically more realistic model of case II.  

According to these results, the motional narrowing of the IR line is indeed 
very similar \cite{oxtoby,lynden} to the motional narrowing of NMR 
resonances \cite{slichter,kubo}.  The motional diminishing of the 
IR-activity of the $g$ mode is, however, an entirely new phenomenon.  
We found that the inverse linewidth of the quasi-static perturbed 
resonance determines the crossover time.  In case I this linewidth is due 
to lifetime broadening, in case II it is determined by the magnitude of 
the perturbation itself.  
Although our discussion was conducted in terms of a classical system, 
experience with quantum oscillators suggests that the main features of this 
crossover behavior would survive a full quantum mechanical treatment.  
All of these phenomena can be treated in a unified way in terms of the 
fluctuation-dissipation theorem \cite{kubo}.

We choose CS$_2$ as a model system to test the predictions of our theory.  
In this material the ``forbidden" Raman lines have been studied previously
in detail as a function of temperature \cite{maddencox,coxmadden,cox} and 
hydrostatic pressure \cite{ikawa,ikawa2,ikawa3,jonas}.  The rotational 
diffusion of the molecules has been 
studied independently by NMR, and the correlation time is well known as
a function of temperature \cite{nmrcs2,nmrcs22}.  

We measured the IR transmission by using a 0.1mm thick liquid cell, 
enclosed by 15mil thick, singly polished, high-resistivity Si windows.  
The cell was filled with Fischer Scientific spectrophotometric 
grade CS$_2$ liquid, sealed 
with teflon washers and mounted on the cold finger of a Helitran liquid 
N$_2$ flow refrigerator.  The spectra were recorded in the 125K-300K 
temperature range by a Bomem MB-155 FT-IR spectrometer, using the spectra 
of the empty cell as backgrounds.  As shown previously, the transmission  
method is well suited for the investigation of weakly-active modes
\cite{c60xtal}.

The spectra around the $\nu_1$ (Raman-active stretching) mode are shown 
in Figure \ref{cs2}.  At 
frequencies extending up to 3000cm$^{-1}$ we clearly observe several 
resonances corresponding to combination modes ($\nu_3-\nu_1$, 
$\nu_1+\nu_2$, $\nu_1+3\nu_2$, $\nu_1+\nu_3$, $2\nu_2+\nu_3$, 
$2\nu_1+\nu_3$, and $\nu_1+2\nu_2+\nu_3$).  
With the sample thickness used in this measurement the IR-active $\nu_2$ 
and $\nu_3$ modes appear totally saturated.  Symmetry considerations 
\cite{nonharmonic,nonharmonic2} on the unperturbed, non-harmonic 
system predict that the combinations composed of an odd number 
of IR modes are IR-active.  The intensity of these modes depends 
on the Bose factors $n_i=n(\hbar \omega _i/ k_{\rm B}T)$ of the 
contributing fundamentals ($i$ is the mode index of the fundamental).  
With the sole exception of the $\nu_3-\nu_1$ difference mode seen in 
Fig. \ref{cs2}, the intensities of the combination modes are expected to be, 
and found to be, approximately independent of the temperature.  
In Figure \ref{tdep} the integrated intensity of this difference mode 
is shown as a function of temperature (circles), together with the expected 
$\vert n_1-n_3\vert$ temperature dependence (dashed line).  

The $\nu_1$ and the $2\nu_2$ resonances are IR-forbidden by the group theory
(they are in fact the two strongest Raman modes).  These lines appear in 
the IR spectrum either due to a symmetry-breaking from the interaction 
between the CS$_2$ molecules or between CS$_2$ molecules and impurities 
in the liquid.  At low temperatures both ``forbidden" lines are
clearly observed.  The $\nu_1$ resonance has a Fano lineshape \cite{fano}.  
As the temperature is increased the lines become broader and their 
intensities decrease, in sharp contrast to the behavior of the ``allowed" 
modes (independent of temperature) and the ``difference" mode (increasing
with temperature).  However, inserting the known NMR correlation times 
into the motional diminishing results described above gives a complete 
account of the temperature dependence (Fig. \ref{tdep}, solid line).  
We emphasize that, other than matching the measured and calculated 
intensities at one temperature, there are no fitting parameters in the 
calculated lines of Fig. \ref{tdep}.  

In a similar manner we can interpret the results of a previous pressure
dependence study of the Raman spectrum of CS$_2$ by Ikawa and Whalley 
\cite{ikawa,ikawa2,ikawa3}.  Increasing pressure in CS$_2$ is analogous to decreasing 
temperature:  the fluctuations slow down ($\tau$ increases).  The Raman 
observation of the IR-active $\nu_2$ mode being sharp at high pressures, 
and broadening and reducing in intensity at lower pressures 
(motional diminishing) can therefore be explained.  Similarly the 
Raman-active polarized spectrum of the $\nu_1$ mode is motionally 
narrowed, while the depolarized $\nu_1$ is motionally diminished at 
lower pressures.

To summarize, using the results of a computer simulated model,
we qualitatively explained the observed diminishing of IR-inactive 
vibrational modes in terms of a fluxuating molecular environment.  
While motional diminishing is not the
sole mechanism which reduces the intensity of weak spectroscopic features
\cite{tabisz2}, we have shown that in appropriate situations its effect 
can be a significant contributor.
The results discussed in this work can be utilized to 
measure fluctuating time scales in solids, liquids and glassy systems.  
In a more 
complex system with IR- and Raman-silent modes, {\em all} weakly active 
fundamentals may exhibit similar motional diminishing given proper symmetry 
breaking fluctuations.  
By investigating the temperature dependence of the 
intensity of these weak modes, molecular fluctuations 
falling into the timescale range determined by the experimentally resolved 
linewidths of spectrometers, typically 300GHz-30THz, can be studied.  

This work was supported by NSF grant DMR9202528.

\begin{figure}
\caption{Time averaged power absorbed as a function of frequency, $\omega$ 
and random fluctuation time, $\tau$, in the frequency range near the $g$
mode.  Inset shows the integrated intensity, its fit to the given functional 
form, and width of the peak as a function of $\tau$.}
\label{gmode}
\end{figure}

\begin{figure}
\caption{Time averaged power absorbed as a function of frequency, $\omega$ 
and random fluctuation time, $\tau$, in the frequency range near the $u$
mode.  Upper inset shows the integrated intensity and width of the peak as 
a function of $\tau$.  Lower inset displays how the character of the 
resonance changes from Lorentzian at low $\tau$ to a mix of Lorentzian and 
Gaussian a high $\tau$.}
\label{umode}
\end{figure}

\begin{figure}
\caption{A portion of the IR transmission spectrum of CS$_2$ as a 
function of temperature.  Labels denote the assignment of the lines,
and the origin of the temperature dependencies.}
\label{cs2}
\end{figure}

\begin{figure}
\caption{Temperature dependencies of the integrated intensity of the 
$\nu_1$ ``forbidden" mode (squares) and the $\nu_3-\nu_1$ difference mode
(circles).  Results from motional diminishing theory is shown as a solid 
line.  The dashed line represents the Bose factor temperature dependence.}
\label{tdep}
\end{figure}


\begin{references}

\bibitem{argon}H.B. Levine and G. Birnbaum, Phys. Rev. Lett. {\bf 20}, 439 
(1968). 

\bibitem{argon2}J.P. McTague and G. Birnbaum, {\it ibid} {\bf 21}, 661 (1968).

\bibitem{sf6}W. Holczer and Y.le Duff, Phys Rev. Lett. {\bf 32}, 205 (1974).

\bibitem{co2}W. Holczer and R. Ouillon, Mol. Phys. {\bf 36}, 817 (1978).

\bibitem{jonas1}J. Jonas and Y.T. Lee, J. Phys. Conens. Matter {\bf 3}, 305 
(1991), and references therein.

\bibitem{birnbaum}``Phenomena Induced by Intermolecular Interactions", Ed. G. 
Birnbaum (Plenum, New York, 1985) .

\bibitem{oxtoby} D.W. Oxtoby, J. Chem. Phys. {\bf 87}, 3028 (1983), and 
references therein.

\bibitem{tabisz2}``Collision and Interaction Induced Spectroscopy" Eds. G.C. 
Tabisz and M.N. Neuman (Kluwer, Dordrech, 1995).

\bibitem{maddencox}P.A. Madden and T.I. Cox, Mol. Phys. {\bf 43}, 287 (1981).

\bibitem{ramansolidcs2}K.D. Bier, H.J. Jodl and A. Loewenhuss, Chem. Phys. 
Lett. {\bf 115}, 34 (1985). 

\bibitem{ramansolidcs22}A. Anderson, {\it et al.}, {\it ibid} {\bf 21}, 
9 (1973).

\bibitem{irsolidcs2}H. Yamada and W.B. Person, J. Chem. Phys. {\bf 40}, 309 
(1964).

\bibitem{irsolidcs22}F. Bertinelli {\it et al.}, Chem. Phys. Lett. {\bf 41}, 
95 (1976).

\bibitem{irsolidc60}C.C. Homes, P.J. Horoyski, M.L.W. Thewalt, and B.P.
Clayman, Phys. Rev. B {\bf 49}, 7052 (1994).

\bibitem{lynden}R.M. Lynden-Bell, Mol. Phys. {\bf 33}, 907 (1977).

\bibitem{slichter}C.P. Slichter, ``Principles of Magnetic Resonance", 
(Harper\&Row, 1963).

\bibitem{oxtoby2}D.W. Oxtoby, Adv. Chem. Phys. {\bf 40}, 1 (1979).

\bibitem{coxmadden}T.I. Cox and P.A. Madden, Mol. Phys. {\bf 39}, 1487 (1980).

\bibitem{winsor}See, for example, P.A. Winsor in ``Liquid Crystals and 
Plastic Crystals" Ed. G.W. Gray and P.A. Winsor, vol 1, p 48 (Wiley, New 
York, 1974).  

\bibitem{c60}W.I.F. David, R.M. Ibberson, T.J.S. Dennis, J.P. Hare, and K.
Prassides, Europhys. Lett. {\bf 18}, 219 (1992). 

\bibitem{c602}J.P. Lu, X.-P. Li, and R.M. Martin, Phys. Rev. Lett. 
{\bf 68}, 1551 (1992). 

\bibitem{c603}P.A. Heiney, J.E. Fischer, A.R. McGhie, W.J. Romanow, A.M. 
Denestein, J.P. McCauley, Jr., and A.B. Smith, Phys. Rev. Lett. {\bf 66}, 
2911 (1991).

\bibitem{kubo}R. Kubo, in ``Fluctuation, Relaxation and Resonance in 
Magnetic Systems", ed. D. Ter Haar, Oliver and Boyd, Edinburgh, p. 23, 1962.

\bibitem{cox}T.I. Cox and P.A. Madden, Chem. Phys. Lett. {\bf 41}, 188 (1976).

\bibitem{ikawa}S. Ikawa and E. Whalley, J. Phys. Chem. {\bf 94}, 7834 (1990).

\bibitem{ikawa2}S. Ikawa and E. Whalley, J. Chem. Phys. {\bf 86}, 1836 (1987).  

\bibitem{ikawa3}S. Ikawa and E. Whalley, J. Chem. Phys. {\bf 85}, 2538 (1986).

\bibitem{jonas}S.L. Wallen, L. Nikiel, J.Yi and J. Jonas, Chem Phys. 
Lett. {\bf 229}, 82 (1994).

\bibitem{nmrcs2}H.W. Spiess {\it et al.}, J. Mag. Reson. {\bf 5}, 101 (1971).

\bibitem{nmrcs22}J.R. Lyerla, D.M Grant, and R.D. Bertrand, J. Phys. Chem. 
{\bf 75}, 3976 (1971).

\bibitem{c60xtal}Michael C. Martin, Xiaoqun Du, John Kwon, and Laszlo Mihaly,
Phys. Rev. B {\bf 50}, 173 (1994).

\bibitem{nonharmonic}See for example ``IR:  Theory and Practice of Infrared 
Spectroscopy", Alpert, Keiser and Szymanski, Plenum Press, pp. 133-137, 1970. 

\bibitem{nonharmonic2}``Introduction to the Theory of Molecular Vibrational 
Spectroscopies", Weedward, Oxford Univ. Press, pp. 329-333, 1972.

\bibitem{fano}This lineshape is well reproduced in our model calculations 
if the asymmetry in restoring force $k^{\prime }$, $k^{\prime \prime}$ is 
accompanied with an asymmetry in the damping $\gamma ^{\prime }$, 
$\gamma^{\prime \prime}$.  For a more detailed discussion, see 
Michael C. Martin and Laszlo Mihaly, {\it to be published}.

\end{references}
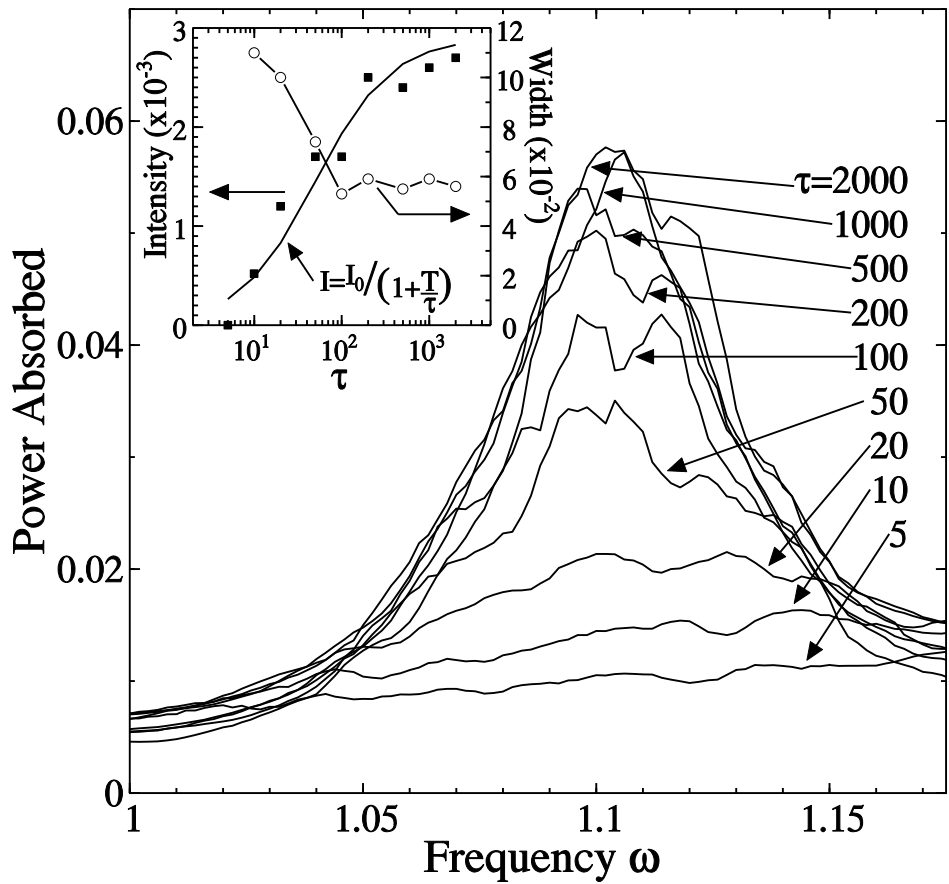
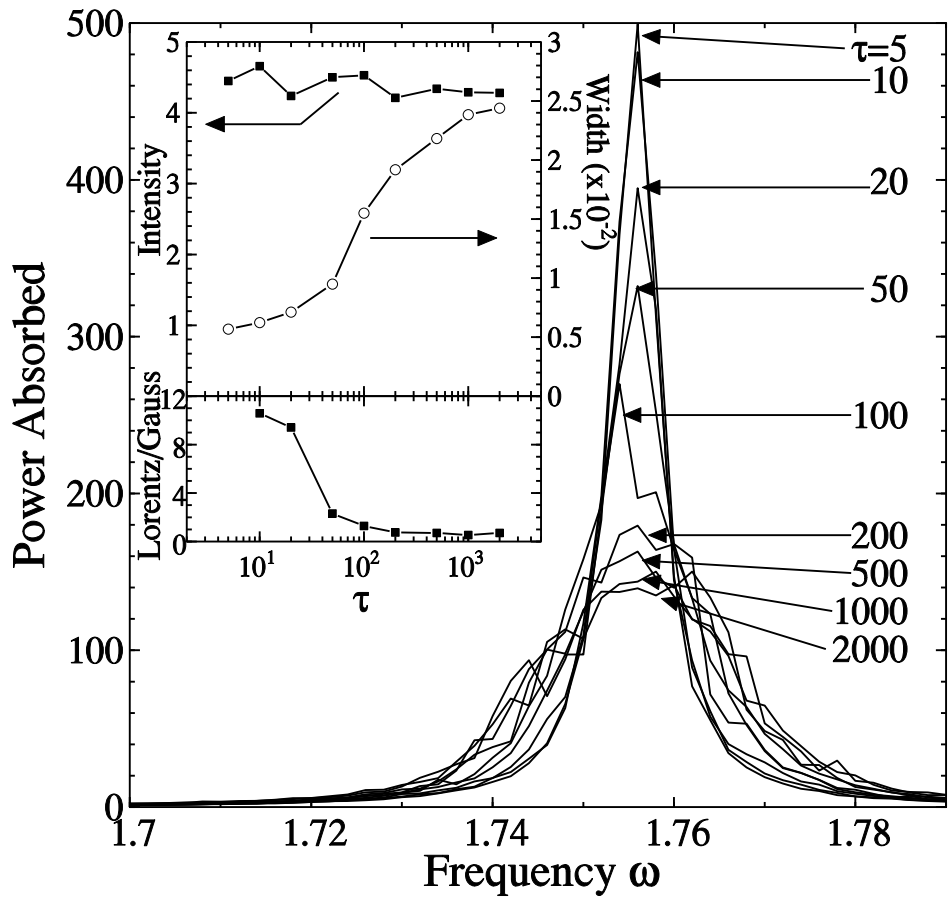
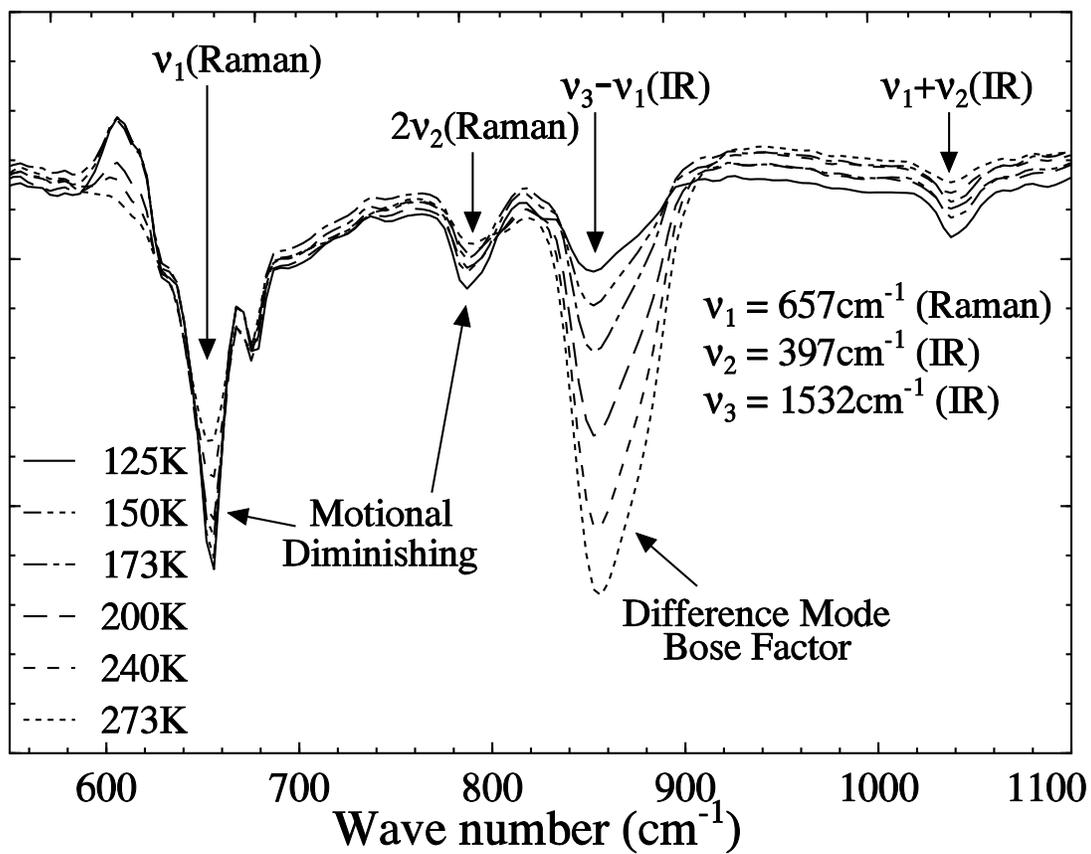
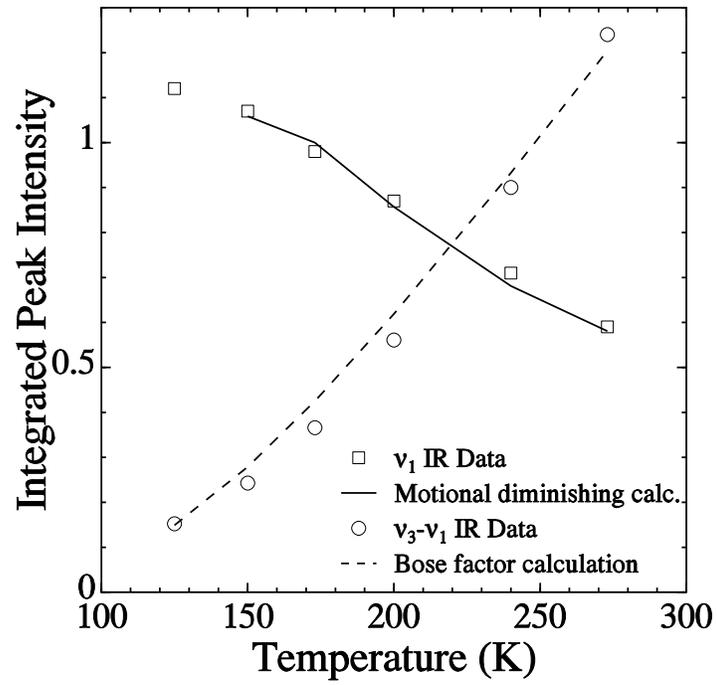
\end{document}